# Quantum interference of surface states in bismuth nanowires probed by the Aharonov-Bohm oscillatory behavior of the magnetoresistance

A. Nikolaeva[1,2]  D. Gitsu,[1]  L. Konopko,[1,2]  M.J. Graf[3] and T. E. Huber.[4]

[1] Academy of Sciences, Chisinau, Moldova.

[2] International Laboratory of High Magnetic Fields and Low Temperatures, Wroclaw, Poland.

[3] Department of Physics, Boston College, Chestnut Hill MA.

[4] Howard University, Washington, DC 20059-0001.

We report the observation of a dependence of the low temperature resistance of individual single-crystal bismuth nanowires on the Aharonov-Bohm phase of the magnetic flux threading the wire. 55 and 75-nm wires were investigated in magnetic fields of up to 14 T. For 55 nm nanowires, longitudinal magnetoresistance periods of 0.8 and 1.6 T that were observed at magnetic fields over 4 T are assigned to $h/2e$ to $h/e$ magnetic flux modulation. The same modes of oscillation were observed in 75-nm wires. The observed effects are consistent with models of the Bi surface where surface states give rise to a significant population of charge carriers of high effective mass that form a highly conducting tube around the nanowire. In the 55-nm nanowires, the Fermi energy of the surface band is estimated to be 15 meV. An interpretation of the magnetoresistance oscillations in terms of a subband structure in the surface states band due to quantum interference in the tube is presented.

PACS.   73.20.-r   71.18.+y    73.23.Ad    73.50.-h

I. INTRODUCTION

The semimetal bismuth has electrons and holes with very low effective masses, and, as a result, electronic quantum confinement effects induce a semimetal-to-semiconductor transformation (SMSC) for wires with diameters below 50 nm which is roughly the Fermi wavelength $\lambda_F$ of bulk bismuth. The SMSC transformation allows control of the band structure through the wire diameter and can be considered as the basis of a method of making materials with attractive properties.[1] For example, theoretical work [2] based on one-dimensional models indicate that Bi wires of small diameters may exhibit superior thermoelectric properties since the density of states at the Fermi level can be tuned, using doping, to very high values and that Bi nanowires may achieve thermoelectric efficiencies of practical interest for wire diameters $d$ less than 10 nm. However, fine nanowires have a large surface area per unit volume and the surface properties of bismuth have to be considered. One important influence of the surface in semiconductors is the appearance of surface states with energy levels in the gap between a valence and conduction band.[3]

Angle-resolved photoemission spectroscopy (ARPES) studies of planar Bi surfaces have shown that it supports surface states, with carrier densities $\Sigma$ of around $5\times 10^{12}\,cm^{-2}$ and large effective masses $m_\Sigma$ of around 0.3.[4] The observed effects are consistent with theories of the surface of non-magnetic conductors whereby Rashba spin-orbit interaction gives rise to a significant population of surface carriers.[5] Measurements of the Fermi surface of small-diameter bismuth nanowires employing the Shubnikov-de Haas (SdH) method [6] have also been interpreted by considering surface charge carriers. Taking into account the surface area per unit volume of these nanowires, there is rough agreement regarding the carrier density and effective mass obtained from measurements of the Fermi surface of 30-nm-diameter Bi nanowires and ARPES measurements. There have been many electronic transport studies of very thin films that hinted at the presence of an excess carrier density on thin films of Bi. [7,8,9] Recently, quantum well states have been detected in very thin Bi films using ARPES[10] indicating that the surface carriers maintain coherence over distances of the order of one nm and that form a discrete number of states as a result. Despite the scientific and technical importance of these effects, experiments that test electronic transport properties of quantum states that appear due to quantum interference in Bi nanostructures are lacking.

Given the bulk electron *n* and hole *p* densities (in un-doped Bi, $n = p = 3 \times 10^{18} cm^{-3}$) and the surface density, $\Sigma = 5 \times 10^{12} cm^{-2}$, measured by ARPES one expects the surface carriers to become a majority in nanowires with diameters of around 60 nm. At that point the nanowire should become effectively a 60-nm tube. The electrical transport properties of tubes of small diameter *d* are unique because the wave-like nature of the charge carriers manifest themselves as a periodic oscillation in the electrical resistance as a function of the enclosed magnetic flux $\Phi=(\pi/4)d^2B$. When the phase coherence length $l_\varphi$ exceeds the circumference, quantum interference between loop trajectories can induce magnetoresistance oscillations with period *h/e* (known as Aharonov-Bohm (AB) oscillations [11]), where *e* is the electron charge and *h* is the Plank constant  The AB phase is $e\Phi/h$. Oscillations with period *h/2e* (known as Altshuler-Aronov-Apivak (AAS) [12]) are also observed. AAS is attributed to interference of a time reverted pair of electron waves and associated with disorder. AB and AAS oscillations have been observed in various conducting rings [13-15], tubes[12] and also in solid cylinders that assimilate tubes such as multiwall-carbon nanotubes (MWNT) [16] and bismuth nanowires in the diameter range 150-1000 nm.[17] In the latter case, theoretical studies have focused on a whispering gallery model of low-effective-mass electrons that define a highly conducting tube in the boundary of the solid cylinder.[18]

Here we report on a study of the magnetoresistance of small-diameter individual Bi nanowires down to 2 K and for magnetic fields up to 14 T. The 55-nm and 75-nm samples that were investigated displayed pronounced $h/2e$ and weak $h/e$ resistance oscillations as a function of magnetic flux at high magnetic fields ($B > 4$ T for 55-nm wires). The observation of these periods is consistent with considering Bi nanowires as a tube of surface states. Since our individual nanowires were long, their contact resistance can be neglected; the elastic scattering length $l_e$ that is found, 250 nm, is longer than the wire perimeter. Since AB quantum interference and ballistic transport, rather than AAS, conditions apply, we are led to an interpretation of the magnetoresistance oscillations in terms of a modulation of the density of states (DOS) which is tantamount to a subband structure of the band of surface states. Deviations of the high field AB period for low magnetic fields parallel to the wirelength and periodic oscillations that are observed for magnetic fields perpendicular to the wirelength are tentatively discussed in terms of spin effects. In sec. II we discuss the experiment and in Sec. III we discuss the results. A summary is presented in Sec. IV.

II. EXPERIMENTS

For this work, samples of individual Bi nanowires were fabricated using a two-step process. In the first step, 200 nm wires were prepared using the Ulitovsky technique, by which a high-frequency induction coil melts a 99.999%-pure Bi boule within a borosilicate glass capsule, simultaneously softening the glass. Glass capillaries (length is in the tens of centimetres) containing the 200-nm Bi filament [17,19] were produced by drawing material from the glass. Encapsulation of the Bi filament in glass protects it from oxidation and mechanical stress. In the second step, the capillaries were stretched to reduce the diameter of the Bi wire from 200 to less that 100 nm. The resultant Bi filament was not continuous, yet sections that are a fraction of a millimeter in length could be selected with the aid of an optical microscope. Electrical connections to the nanowires were performed using $In_{0.5}Ga_{0.5}$ eutectic. This type of solder consistently makes good contacts, as compared to other low-melting-point solders, but it has the disadvantage that it diffuses at room temperature into the Bi nanowire rather quickly. Consequently, the nanowire has to be used in the low temperature experiment immediately after the solder is applied; otherwise the oscillations of the magnetoresistance that we discuss are not observed. The samples used in this work are, to date, the smallest diameter single Bi wires for which the electronic transport at low temperatures has been reported. The wires are characterized by an electronic diameter $d = \sqrt{\rho_{Bi} l / R_o}$, where $\rho_{Bi} = 1.1 \times 10^{-4}$ Ω-cm is the 300 K bulk resistivity, $l$ is the wire length, and $R_0$ is the wire resistance. In the present study, we used samples with $d=$ 55 and 75 nm. A scanning electron microscope (SEM) image of the 55-nm nanowire is shown in the inset of Fig. 1.

In addition to the large surface area per unit volume, quantum confinement also plays a role in making the surface carriers dominate the electronic transport in Bi nanowires of small diameter. For T>100 K the nanowires' resistance $R(T) \sim \exp(\Delta/2k_B T)$, where $\Delta$ is interpreted as the energy gap between the electron and hole band in the core of the nanowires. $\Delta$ is 14 meV for the 55-nm wire. The value of the gap that is observed is in rough agreement with theoretical estimates that put the critical diameter for the SMSC for binary wires at around 50 nm.[2] Accordingly, at low temperatures when the thermal energy is small (i.e. $k_B T$ =0.13 meV when $T$ = 1.5 K), the bulk states do not contribute to electronic transport significantly and it can be surmised that the low temperature electronic transport that is observed is mediated by surface states. The square-resistance $R$ of surface states is, therefore, found to be around 300 Ω.

$B$-dependent resistance ($R(B)$) measurements in the 0 to 14 T range were carried out at the International High Magnetic Field and Low Temperatures Laboratory (Wroclaw, Poland) and we employed a device that tilts the sample axis with respect to the magnetic field and also rotates the sample around its axis; the angles are defined in Fig. 2, inset. Figure 2 shows the 1.5 K $R(B)$ of a 55-nm wire for $\alpha = 0°$ and for a case where $\alpha =90°$. The former and latter cases are the longitudinal magnetoresistance (LMR) and transverse magnetoresistance (TMR) cases, respectively. In the LMR case, $R(B)$ decreases for increasing magnetic field and at 14 T, $R(B)$ is ~100 KΩ, which is an estimate of the upper value of the contact resistance. The decrease of the resistance with applied magnetic field parallel to the wirelength is typical of Bi nanowires of large diameter nanowires [19] and will be discussed further below. Although conductance fluctuations were observed in Bi nanowires[20] and thin Bi film quantum dots, [21] they are not observable for the nanowires in the present study, presumably because the nanowires are long. A search for Shubnikov-de Haas oscillations (periodic in $1/B$) in these samples was unsuccessful; SdH oscillations are observed in the 200-nm Bi nanowires from where we obtained the 55-nm nanowires, showing that the bismuth material in the nanowires has the required high purity and high mobility. SdH oscillations have been observed in 30-nm Bi nanowire arrays;[6] however, in the samples in the present study, the conditions for observing SdH are not optimal due to the large AB oscillations. As shown in Fig. 3, the resistance changes when the nanowire is rotated around its axis. The rotational diagram of this 55-nm nanowire is similar that the one shown by large diameter nanowires, with minima and maxima separated by 90 degrees. It has been observed that large diameter individual nanowires are single-crystals whose crystalline structure, as determined by Laue X-ray diffraction and SdH methods, is oriented with the ($10\bar{1}\bar{1}$) along the wire axis.[19]. In such nanowires the resistance minima and maxima correspond to the magnetic field aligned with the *C2* and *C3* axis. Therefore it is reasonable to assume that the nanowires in the present study are also single-crystals.

Figure 4 shows *dR/dB* for various orientations of the magnetic field. A large modulation of the resistance is observed for all angles $\alpha$, decreasing in intensity for increasing magnetic fields. We have preliminarily studied other 55-nm and 75-nm samples, and they exhibit similar modulation of the resistance so that the effect appears to be a property of all Bi nanowires in this range of diameters. For low magnetic fields, *B* < 4 T, the modulation, which is 50 kΩ, is comparable to the characteristic quantum resistance $h/e^2 \approx 25.8$ kΩ. Figure 5 shows extrema positions. The oscillation's periodicity is sustained, as the extrema lie on straight lines over extended ranges of magnetic fields (low-*B*, intermediate and high-*B*), indicated by the dashed delimiting lines. Figure 6 shows the period *P* as a function of orientation in the low-*B* (0-3.5 T) and high-*B* (4.5-14 T) magnetic field ranges in the LMR case. In addition, figure 6 shows the orientation-dependent periods observed for 560-nm Bi nanowires from reference 17. The high-*B* periods exhibit the angular dependence $P \sim (\cos^{-1}(\alpha))$ that is expected if the interference occurs in a path perpendicular to the wirelength regardless of the orientation of the magnetic field since, in this case, $\Phi = (\pi d^2/4) B \cos\alpha$. This is the angular dependence that has been observed in large diameter Bi nanowires [17] and also in MWNT.[22] At high *B*, the oscillation period of the 55-nm wire can be fit with P = 0.79 T $\cos^{-1}(\alpha)$; the $\alpha = 0$ period is in fair agreement with $2h/(e\pi d^2)$ = 0.85 T, as calculated from the diameter of the wire by assuming a flux period of *h/2e*. In contrast, at low-*B*, oscillation periods are largely $\alpha$-independent. For $\alpha < 20°$, the high-*B* oscillation periods are shorter than the corresponding low-*B* periods by a factor of 1.6 $\pm$ 0.1. Also, at low-B, we find that the TMR presents oscillations; such oscillations are not found in the high-B range. Study of the periodicity of *dR/dB* employing Fourier transform analysis in the high-*B* range yields yet another mode, a long oscillation period that can be fit as 1.58 T $\cos^{-1}(\alpha)$, and is therefore consistent with a flux period of *h/e*. Finding the *h/e* and the *h/2e* modes together is not surprising; this has been observed in rings where it has been interpreted as been due to the interference of electrons that encircle the ring twice.[13]

Figure 7 shows the magnetic field derivative of the longitudinal magnetoresistance of 75-nm nanowires. The inset shows maxima positions versus B. The high-B period is 0.48 T. Within the experimental errors, the high–B period is proportional to the inverse of the wire cross-sectional area, confirming our interpretation that the high-field oscillations result from the AB effect.

Figure 8 shows the *dLMR/dB* at various temperatures. The oscillations are attenuated substantially as *T* increases from 1.5 K to 4.8 K. We attempted to differentiate the oscillations of long period at low-B from the ones of short period at high-B according to the temperature dependence of the corresponding amplitudes, but we observed only small differences in temperature-dependent behaviour, suggesting that the low-$B$ and high-$B$ oscillations are not different modes.

III. DISCUSSION

For the sample with $d = 55$ nm, our measurements show that the bulk electron and hole bands are separated by a gap of 14 meV. We have not discussed the band of surface states. We assume a parabolic dispersion relation which is consistent with reference 4 and also with the observation of a simple arrangement of Landau states (periodic in 1/B).[6] The number of states (NOS) per spin state of a 2D system is $m_0 m_\Sigma E_F / \pi \hbar^2$ where $m_0$ is the free electron mass and $E_F$ is the Fermi energy. Taking $m_\Sigma = 0.3$, for $\Sigma = 2 \times 10^{12} \, cm^{-2}$ (these values were measured using SdH in Ref. 6 for 30-nm Bi nanowires), we find that $E_F = 18$ meV for 55-nm nanowires, which is in fair agreement the value of the gap (14 meV) observed. For $\Sigma = 5 \times 10^{12} \, cm^{-2}$, which is the value of the surface density according to reference 4, we would find that $E_F = 39$ meV, which is not consistent with our observation of a gap of 14 meV between the electron and hole bands (the surface band would not fit between the top of the T-hole band and the bottom of the L-electron band). The schematic energy band diagram showing the energies of the band edges and the Fermi level for bulk Bi and the 55-nm Bi nanowire is presented in Fig. 9.

For the nanowires in the present study we can estimate the mean free path of the surface carriers as follows. The density of states $\mathcal{D}(E)$ is $m_0 m_\Sigma / \pi \hbar^2$ and therefore the diffusion coefficient $D = 1/(R\ e^2 \mathcal{D}(E_F))$ is 208 cm$^2$/sec. This diffusion coefficient is high in comparison with those discussed by Aronov and Sharvin [12] for hollow cylinders and low in comparison with Bi thin films. [21] Since the Fermi velocity $V_F$ is $5 \times 10^6\, cm/\sec$, the elastic scattering length $l_e$ is $3D/V_F$ =250 nm which is somewhat longer than the circumference of the nanowire.

We discuss the oscillation of the magnetoresistance as evidence of a subband structure in the surface carriers band as follows. We sought guidance in other cases of AB phenomena in solid state samples such as antidot arrays [23] and in nanotubes where the condition for quantum interference is met. In tubes this condition is $l_e >$ perimeter. In single-wall carbon nanotubes, penetration by an entire flux quantum requires inaccessibly high magnetic fields. In MWNT the AB oscillations that are observed [24] (tubes are shorter than the dephasing length in order to inhibit AAS phenomena) are attributed to structure of the two-dimensional density-of-states (DOS). Each peak of the DOS is due to the opening of a 1D channel of conduction in the wire associated with a particular chiral state.[25] The chiral states in carbon nanotubes are the sp states discussed by Ajiki and Ando [25] for graphene surfaces. Considering the similarities between the Bi surface states tube and carbon tubes, we propose an interpretation of the oscillations of the magnetoresistance that we observe in terms of oscillations in the density of surface states. The energy of surface carriers in the tube is the sum of the 1D translational energy and the orbital energy that includes spin. Since the density of states of the 1D system is a maximum for zero energy, then the oscillations of the MR indicate magnetic fields for which the orbital energy is equal to the Fermi energy. The quantum mechanical problem of a particle in a circular path in the presence of Rashba spin orbit and Zeeman coupling has been studied extensively in the case of magnetic field perpendicular to the circle which corresponds to our LMR case. We adopt the solution presented by Nitta, Meijer and Takayanagi (NMT) [26] The strength of the spin-orbit interaction is given by $B_{SO}$. The orbital energy depends on orbital quantum number $m$, the rotation direction $\lambda$ and the spin direction $\mu$. $m$ is an integer. $\lambda$ is +1 for clockwise rotation and −1 for counterclockwise rotation. $\mu$ is +1 and −1 for spin-up and spin-down, respectively. For large $m$'s and for large magnetic fields ($B > B_{SO}$), the energy of the orbital states is:

$$E_\mu^\lambda = \frac{2\hbar^2}{md^2}(m + \lambda\frac{h\Phi}{c})^2 + \mu\frac{\hbar geB}{4mc} \tag{1}$$

The first and second terms are the kinetic and Zeeman energies, respectively. Here, g is the electron g-factor that is two for free electrons and $d$ is the tube diameter. This the proposed equation for the energy levels of the subband structure in the surface states band in the LMR case. We will discuss below that, with small modifications, expression [1] is valid for $B < B_{SO}$ also. Evaluating expression [1] for B=0, we find $m = 12$. As we increase the magnetic field, we find states of decreasing $m$ and various rotational directions and spin directions with a maximum number of four levels for a change of magnetic field corresponding to $h/e$ which is 1.5 T. Like those quantum well states observed in Ref. 10, these states are due to quantum interference with the difference being that the geometry in [10] is planar and in our case is cylindrical. Another difference is that our states are composed of an orbital wavefunction that is coupled to a one-dimensional conduction channel. There is a discrepancy between the model that arises from expression 1, that gives four level crossings per $h/e$ period and the observation of two level crossings per $h/e$ period that is observed. This discrepancy is solved below.

There are a number of effects beyond the scope of Equation 1 that we have observed in the low-B range and are not observed in the high-B range in the nanowires. One effect is a gradual increase of the period for decreasing magnetic field in the LMR case. This is shown in Fig. 5 for 55-nm and in the inset of Fig. 7 for 75-nm nanowires. The crossover field is 4 T and 3 T for 55-nm and 75-nm nanowires, respectively. We also observe that, in the low-B range, the TMR displays an oscillating magnetoresistance; in other words, there is an oscillation associated with a change of magnetic field under conditions where the flux through the cylinder $\Phi = 0$. Rashba spin-orbit coupling (SOC) in the surface states may explain these observations. Rings of 2D holes in GaAs have been studied extensively and, similar to our nanowires, show AB oscillations and involve SOC. In this case, it is observed that the magnetoresistance shows in addition to the fundamental oscillation that is cyclic with the AB phase, a sideband, whose strength depends upon the range under observation.[14] The sideband is interpreted as evidence of a Berry or dynamical phase. The phase originates from the change in the direction of the local field which is the resultant of $B$ and the effective SOC field $B_{SO}$. Since the effects due to the Berry phase appear for $B < B_{SO}$ it is natural to associate the effective SOC field with magnetic fields (4 T for 55 nm nanowires) for which we observe a crossover from low-B to high-B behavior. However, the conditions for observing Berry phase phenomena are very stringent [15] and the observations in Ref. 14 are considered controversial. A more in-depth study of the experimental results concerning spin-orbit coupling in Bi nanowires will be presented in a separate publication.

We observed a decrease of the resistance with an applied magnetic field parallel to the wirelength (figure 2). This effect has been observed in many studies and by many groups in almost all samples of Bi nanowires even those of small diameter.[2,27] When conduction is dominated by L-electrons this phenomenon, Chamber's effect, occurs when the magnetic field focuses electrons towards the core of the wire (away from the surface) thereby avoiding surface collisions. In the case of bulk conduction, Chambers effect is associated with ballistic transport. However, we have argued that electronic transport is dominated by surface carriers and the question that arises is the interpretation of the decrease of resistance with magnetic fields for surface carriers. We propose that it is conceivable that surface carriers of the rotation direction $\lambda$ = +1 experience decreasing surface scattering for increasing magnetic fields because of magnetic forces that point inward. Since this force depends upon the rotation direction, charges with $\lambda = -1$ would experience more surface scattering which preclude them from giving observable peaks in the magnetoresistance. Since there are two of such states per $h/e$ period, not four, this interpretation would resolve the discrepancy about the number of level crossings. Also, it would demonstrate spin-splitting because the states of different spin give rise to separate peaks of the DOS. We are working on a detailed model of surface scattering based on the consideration of lateral confinement of surface states and preliminary results are encouraging.

Now we comment on the experimental observation that the prominent oscillations in 55 nm nanowires are AB and are SdH in 30 nm nanowires. Surface states are localized at the surface; inside the crystal the wave functions show damped oscillations with increasing distance from the surface. The observation of SdH oscillations due to surface states in 30 nm nanowires in reference 6 is significant. SdH oscillations are caused by Landau states near the Fermi level. Landau states are due to the quantization of closed orbits in the presence of magnetic fields; their orbits extend over the Larmor radius $r_L$ which is $m^* m_0 V_F / |e|B$. For B = 5 T, we get $r_L$ = 17 nm. To support Landau states, the spatial range of the surface states has to be larger than $r_L$. Therefore the surface states in 30 nm nanowires fill a very significant fraction of the wire and it cannot be considered a hollow cylindrical conductor. This explains why AB phenomena cannot be easily identified in 30-nm nanowires and we reported only the observation of SdH at high magnetic fields. Also, in comparison to wire arrays, the conditions for observation of quantum interference effect are vastly improved in individual nanowires; in arrays of nanowires the diameter are not entirely uniform, the phase information relevant for quantum interference is averaged, and therefore the magnetic field for maxima and minima randomized. Further work regarding the magnetoresistance of 30 nm arrays at low magnetic fields is in progress.

We note that since the surface carriers have high (metallic) density and their effective masses are high, their partial thermopower is much smaller than that of semimetallic Bi; still, the surface states may contribute to the thermopower of Bi nanowires significantly due to the presence of a subband structure in the surface carriers band and an associated structure in the density of states.

IV. SUMMARY

In conclusion, we present a band model for small diameter Bi nanowires that accounts for our resistance and magnetoresistance measurements. In this model, at low temperatures, the L-electrons and T-holes states are empty and the charge carriers are predominantly in surface states in the periphery of the nanowire. Charge carriers are confined to the hollow conducting cylinder made of surface states. When this conducting tube is threaded by the magnetic field, surface carriers wave-like nature manifests itself as a periodic oscillation in the electrical resistance as a function of the enclosed magnetic flux. The oscillation of the magnetoresistance is revealing of a subband structure in the band of surface carriers. This subband is due to the presence of orbital states under quantum interference conditions that are similar to the quantum well states observed using ARPES in very thin Bi films. Effects at small magnetic fields that are expected for spin-orbit Rashba interactions are observed and therefore, our study may provide the basis for exploring spin-dependent transport in nanowires.  Measurement of the properties of surface charges is an integral part of the problem of quantum size effects and thermoelectricity of Bi nanowires.

The authors wish to thank Y. Hill and P. Jones for useful discussions. This work was supported by the Civilian Research and Development Foundation for the Independent States of the Former Soviet Union (CRDF), award #MP2-3019. T.E.H. and M.G. work was supported by the Division of Materials Research of the U.S. National Science Foundation under Grants No. NSF-0506842 and NSF-0611595. T.E.H. work was also supported by the Division of Materials of the U.S. Army Research Office under Grant No. DAAD4006-MS-SAH.

**FIGURE CAPTIONS**

Figure 1. Temperature dependent resistances of 55- and 75-nm diameter single Bi nanowires. The length of the 55-nm nanowire is 0.65 mm. The resistance of the 75-nm wire has being scaled to match the 55 nm data at 300 K. Inset: SEM cross-sectional image of the 55-nm wire (clear) in its 20-μm glass envelope (gray background). SEM cross-sections are obtained by polishing that may grind the glass preferentially, making the nanowire appear larger than what it is; in the SEM in the figure, the nanowire cross-section measures 70 nm.

Figure 2. Resistance of the 55-nm wire at 1.5 K with an applied magnetic field. The angle between the magnetic field and the wirelength is indicated.

Figure 3. Rotational diagrams of the 75-nm nanowire showing the dependence of the value of resistance at 4 K for fields of constant value and perpendicular to the wirelength ($\alpha=90^{\circ}$) as a function of $\beta$. $C_2$ and $C_3$ indicate the $\beta$'s for which $\vec{B}$ is aligned with the binary and trigonal axis of the tetragonal system.

Figure 4. $dR/dB$ for the 55-nm wire at 1.5 K for various orientations of the applied magnetic field with respect to the wirelength. $\beta=0^{\circ}$. The curves have been displaced vertically for clarity.

Figure 5.   Indexing of the assigned position of maxima (integer) and minima (half-integer). The solid lines are linear fits. The dotted lines are an aid to the eye. The dashed lines delimit *B*-ranges discussed in the text.

Figure 6.   Solid circles: Angle-dependent periods obtained from the magnetoresistance data in Fig. 5. Empty rectangles: Additional periods obtained via Fourier transform analysis; the rectangle height indicates the experimental error. Empty circles: AB oscillation periods reported in Ref. 17 for 560-nm Bi nanowires. The dashed line is a fit to that data with period $\Delta = 0.018$ T/cos($\alpha$). Inset: Fourier transform of the resistance in the $\alpha = 0°$ case, in the interval 8-14 T, showing peaks at 0.89 and 1.45 T.

Figure 7.   *dR/dB* for the 75-nm wire at 1.5 K for $\alpha=0$. Inset: Indexing of maxima versus B.

Figure 8.   *dR/dB* ($a = 0$) of 55-nm wires for three temperatures as indicated. The inset shows the amplitude's temperature dependence for the first ($B \sim 1.5$ T) and fourth ($B \sim 4.1$ T) maxima.

Figure 9.   Schematic energy band diagram showing the energies of the band edges for the *L*-point electron pockets and the T-point holes and the Fermi level for: (a) bulk Bi, where the band overlap is 38 meV and the Fermi level is 27 meV from the bottom of the *L*-point pocket band edge, and (b) 55 nm Bi nanowires where the band gap $\Delta$ is 14 meV and $E_F \sim \Delta$. The surface states band is shown with dashed lines

to highlight the fact that the density of surface per Bi atom is significantly less than that of bulk electrons and holes.

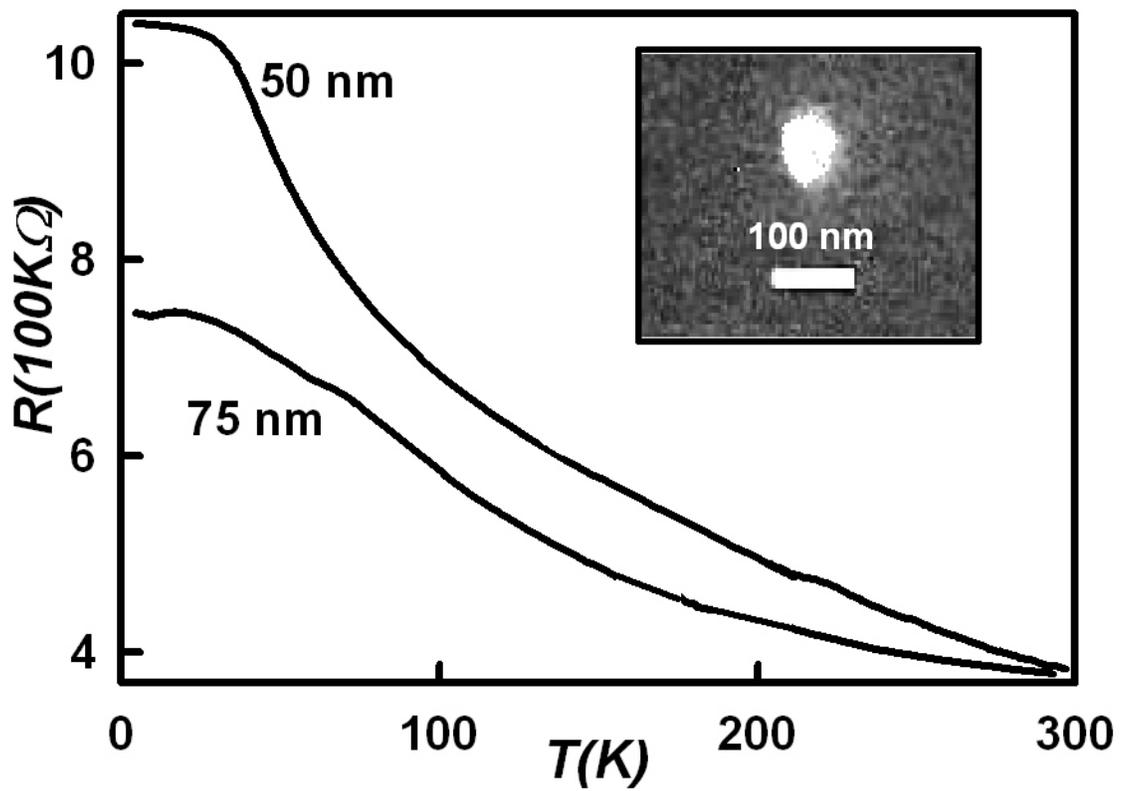

Figure 1. Nikolaeva *et al.* (2007).

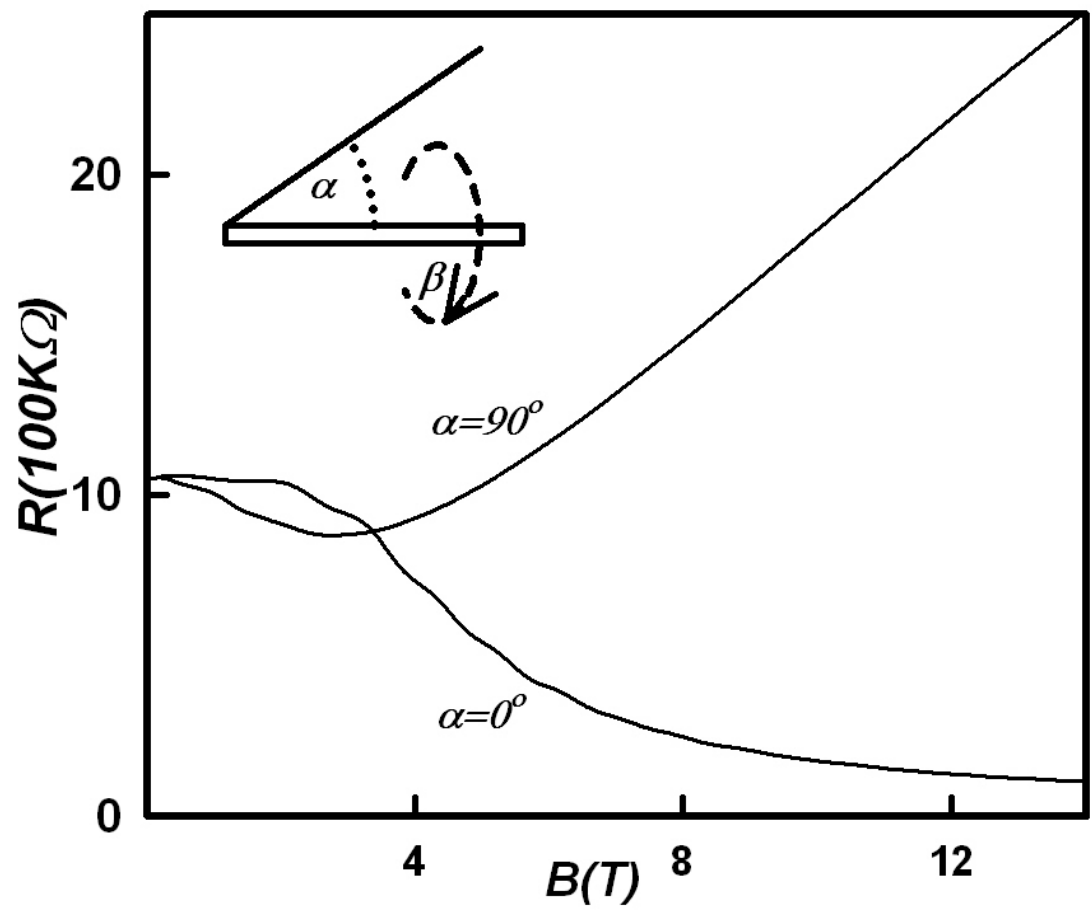

Figure 2. Nikolaeva *et al*. (2007)

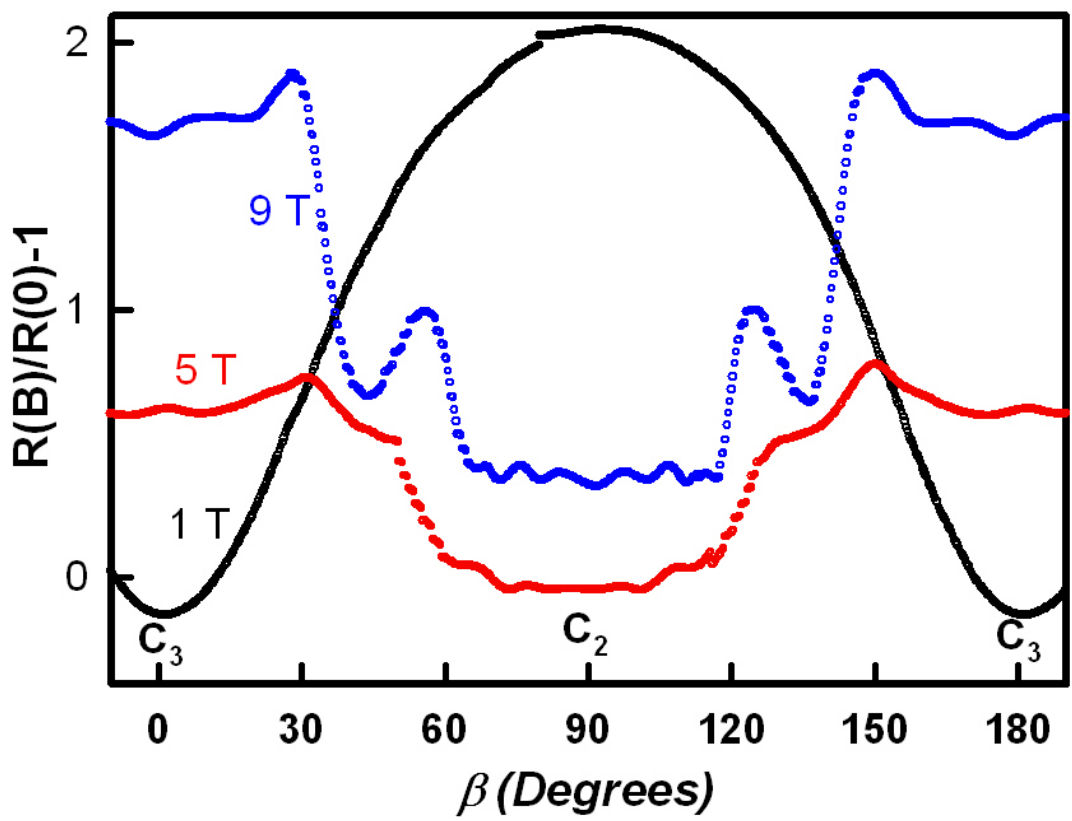

Figure 3. Nikolaeva *et al.* (2007)

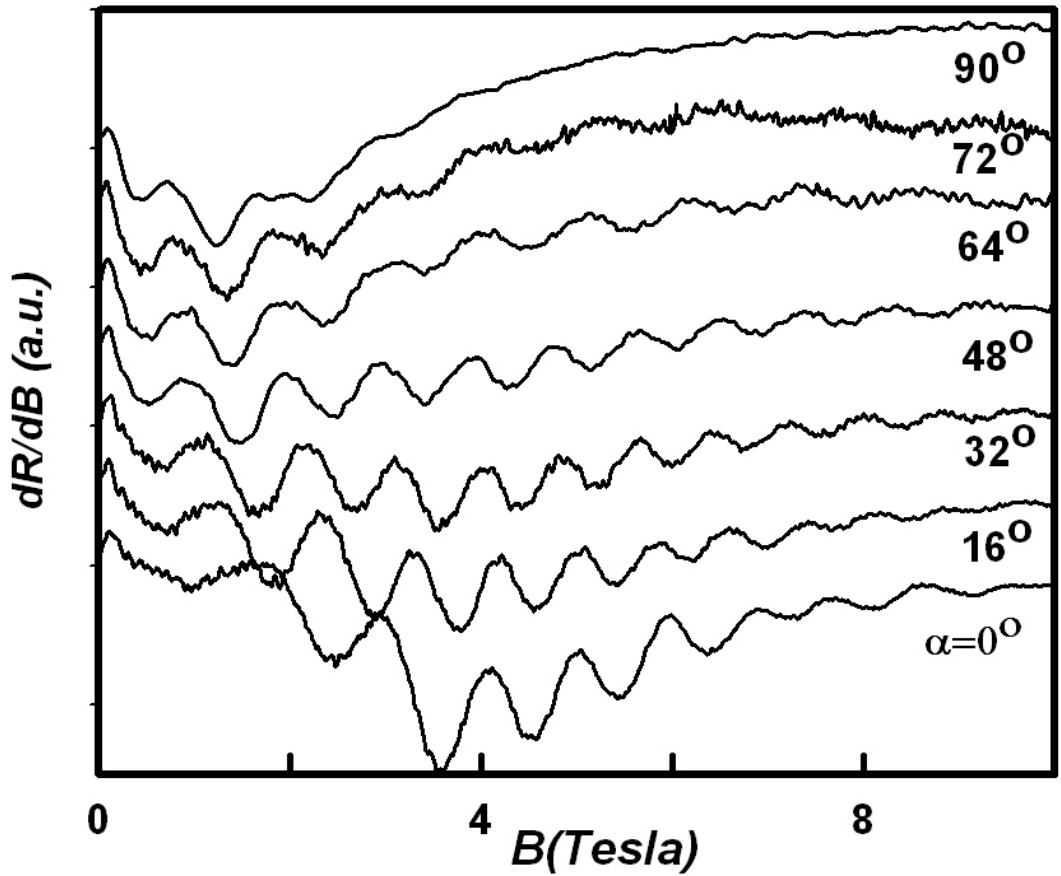

Figure 4. Nikolaeva *et al*. (2007).

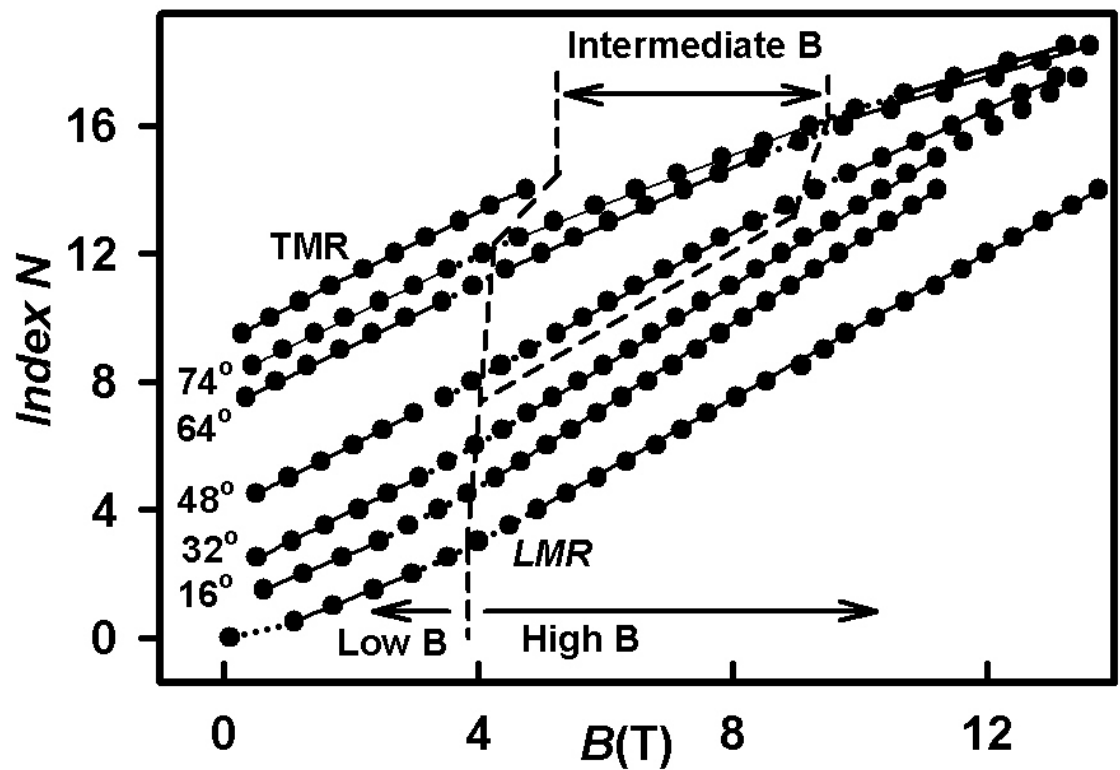

Figure 5. Nikolaeva *et al.* (2007).

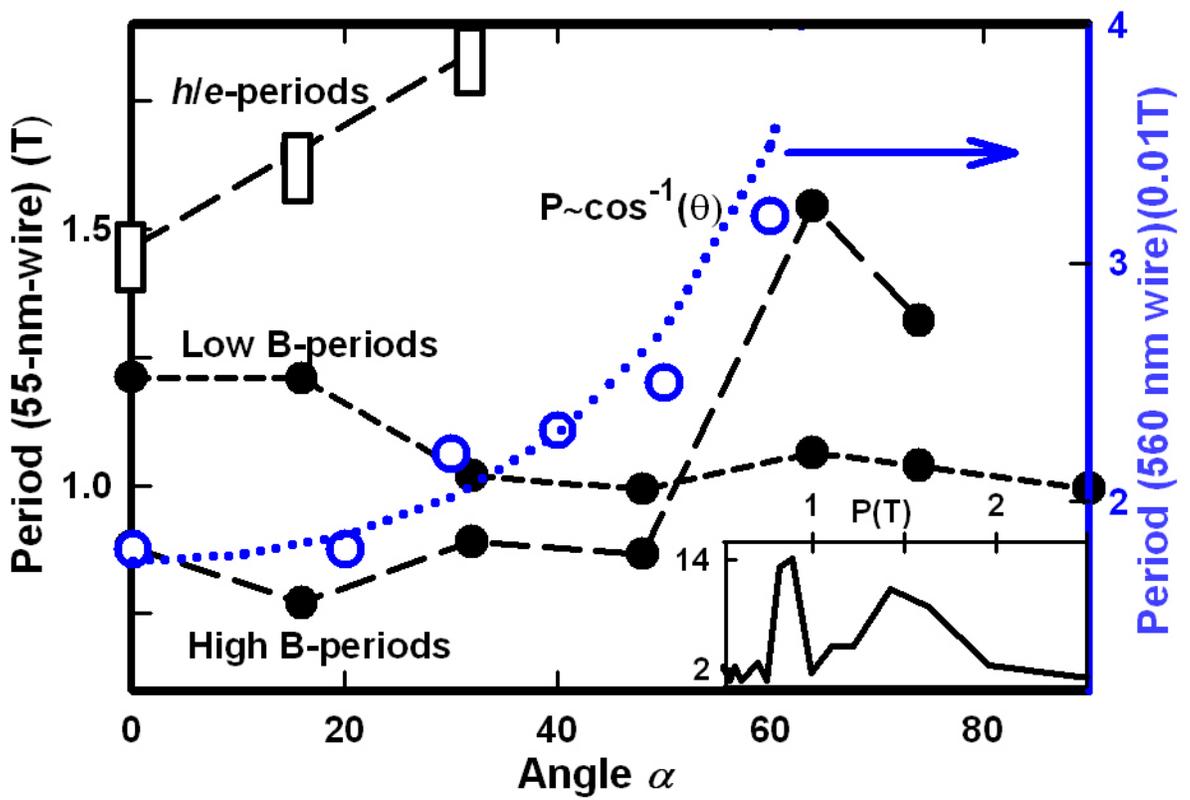

Figure 6. Nikolaeva *et al.* (2007).

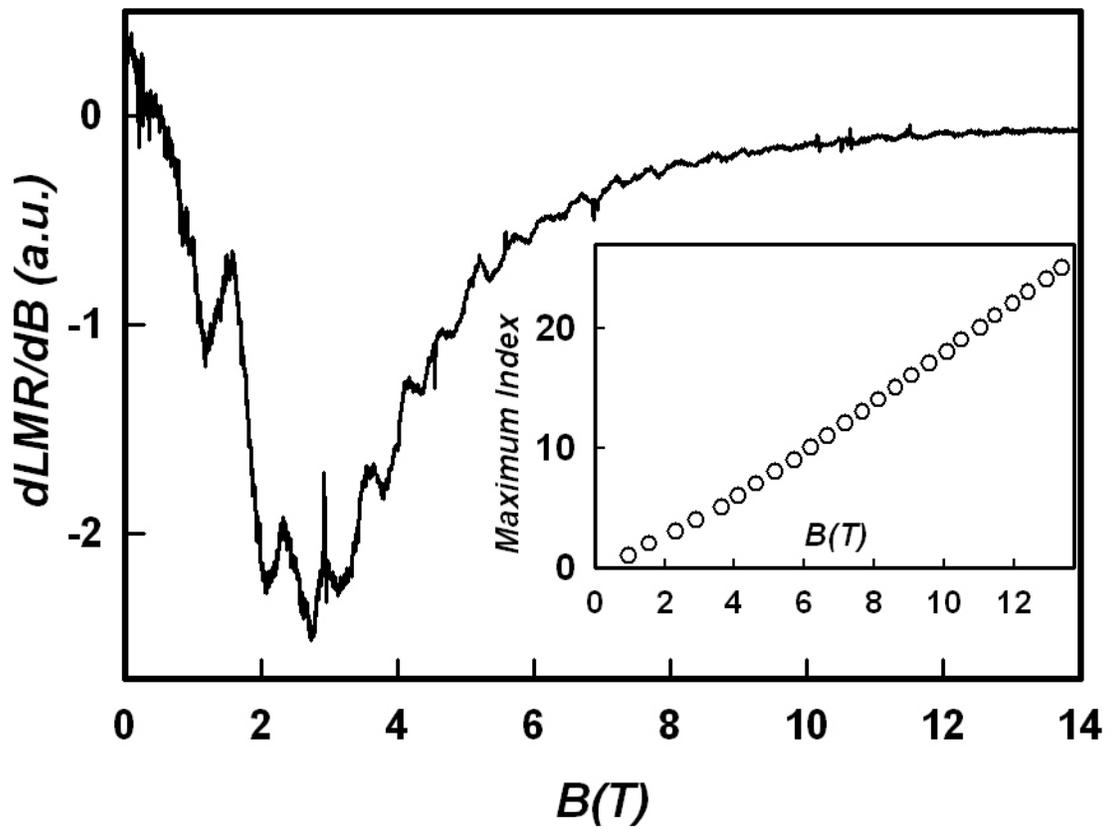

Figure 7. Nikolaeva *et al*. (2007).

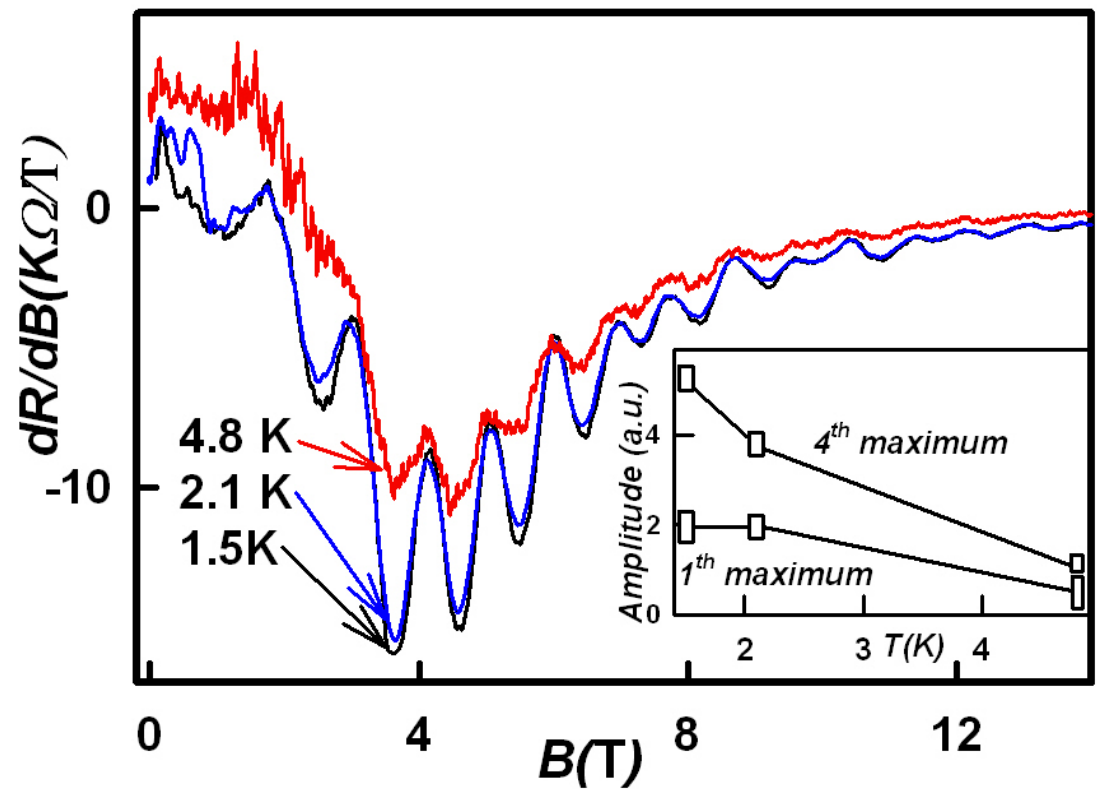

Figure 8. Nikolaeva *et* al (2007).

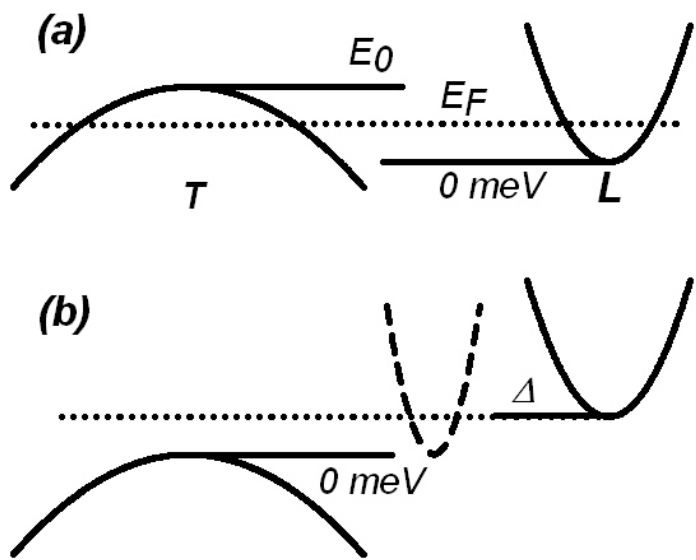

Figure 9. Nikolaeva. (2007).